\documentstyle[12pt]{article}

\begin{document}
\begin{titlepage}
\begin{flushright}
UTTG-08-96 \\
hep-th/9606133 \\
June, 1996
\end{flushright}
\vspace*{2cm}
\begin{center}
{\LARGE\bf F-Theory Duals of Nonperturbative \\ 
       Heterotic $E_8\times E_8$ Vacua in Six Dimensions}
\end{center}
\bigskip
\begin{center}
{\large Philip Candelas\footnote[1]{candelas@physics.utexas.edu}, Eugene 
        Perevalov\footnote[2]{pereval@physics.utexas.edu} and Govindan 
        Rajesh\footnote[3]{rajesh@physics.utexas.edu}}
\end{center}
\bigskip
\begin{center}
        {\it Theory Group \\ Department of Physics \\ University of Texas \\
           Austin, TX 78712, USA}  
\end{center}
\bigskip
\begin{abstract}
 We present a systematic way of generating F-theory models dual 
to nonperturbative vacua (i.e., vacua with extra tensor multiplets) of 
heterotic $E_8\times E_8$ strings compactified on $K3$, using hypersurfaces
in toric varieties. In all cases, the Calabi-Yau is an elliptic fibration
over a blow up of the Hirzebruch surface {\rm I\kern-.18em F}$_{n}$.
We find that in most cases the fan of the base of the elliptic fibration is 
visible in the dual polyhedron of the Calabi-Yau, and that the extra tensor
multiplets are represented as points corresponding to the blow-ups of the
{\rm I\kern-.18em F}$_{n}$. 

\end{abstract}
\end{titlepage}

{\bf Contents}
\bigskip
\begin{enumerate}
\item Introduction
\item F-theory/Heterotic Duality in 6 Dimensions
\item Nonperturbative Vacua and Reflexive Polyhedra
  \begin{enumerate}
   \item Blowing up the Base
   \item Reflexive Polyhedra And Hodge Numbers
   \item Unhiggsing $E_8$
  \end{enumerate}
\item Conclusion 
\end{enumerate}
\pagebreak

\section{Introduction}
In this paper we study six dimensional, $N=1$, theories arising from
compactifications of F-theory on elliptic Calabi-Yau  threefolds. Such 
theories are conjectured to be dual to heterotic $E_8\times E_8$ strings
compactified on $K3$ \cite{MV2}. Upon further compactification on a torus,
this yields \cite{V:F} the duality between heterotic $E_8\times E_8$ 
strings on 
$K3\times T^2$ and the type IIA string on the same elliptic Calabi-Yau
 \cite{KV,FHSV}. This duality has been the subject of extensive study in 
the recent literature [3 - 14].

The cases studied in \cite{MV2} were duals to perturbative heterotic vacua
on $K3$ with instanton numbers $(12 - n, 12 + n)$ for the $E_8$'s. Also, it
was shown that blowing up the base of the elliptic fibrations produced new
models with additional tensor multiplets in six dimensions.

The purpose of this paper is to present a systematic method, 
using hypersurfaces in 
toric varieties, of constructing elliptic 
Calabi-Yau threefolds which provide the F-theory duals of nonpertubative
heterotic vacua with arbitrary numbers of tensor multiplets in six dimensions  
subject to the anomaly cancellation condition \cite{DMW} 
\[ d_1 + d_2 + n_T = 25\] 
(where
$d_1$ and $d_2$ are the instanton numbers in the two $E_8$'s 
respectively, and $n_T$ 
is the number of tensor multiplets). Our method generalises the ideas in 
\cite{MV2}. We show that there is a simple relation between the distibution of 
instantons and tensor multiplets on the heterotic side and the  
Weierstrass equation describing the Calabi-Yau manifold on the F-theory side.
Our method differs from the procedure outlined in
\cite{poly:CF,LSTY} in that we specify the action of the various 
{\bf C}$^\star$'s on the 
homogeneous coordinates, instead of adding points to the dual polyhedra of the 
Calabi-Yau 
manifolds by hand.  
In all cases, we find that the Hodge numbers
calculated using Batyrev's toric approach \cite{B, COK} are 
consistent with the
results on the heterotic side. Furthermore, we find that the dual polyhedra of
these manifolds exhibit the structure described in \cite{poly:CF,LSTY}.
We also study Calabi-Yau threefolds describing the F-theory duals of heterotic
vacua with an enhanced $E_8$ gauge symmetry. In such a situation, all the
instantons in that $E_8$ are necessarily small - smoothing out the singularity
requires blowing up the base, so that we get as many extra tensors as
we had instantons to start with \cite{PAsp}.

This paper is organized as follows. In section 2 we give a brief review of 
F-theory/heterotic duality. In section 3 we describe our procedure for 
constructing the Calabi-Yau threefolds providing F-theory duals of 
heterotic vacua
with tensor multiplets and arbitrary instanton numbers in the two $E_8$'s
(subject always to the anomaly cancellation condition mentioned above). 
We study
the dual polyhedra of these manifolds and compare them with the polyhedra 
studied in 
\cite{poly:CF}. We also calculate the Hodge numbers using the methods of 
Batyrev
\cite{B, COK} and find agreement with the results of calculations on the 
heterotic side. Next, we describe our procedure for constructing duals 
of heterotic models with enhanced $E_8$ gauge symmetry.
Finally, section 4 summarises our results.

\section{F-theory/Heterotic Duality \\ in 6 Dimensions}
In this section we give a brief review of the F-theory/heterotic duality
in 6 dimensions mainly following \cite{MV2}. It was shown in \cite{V:F}
that heterotic string theory compactified to 8 dimensions on $T^2$ is dual
to the F-theory compactified on $K3$ which admit an elliptic fibration.
The elliptic fiber can be represented by the Weierstrass equation:
\[ y^2=x^3+ax+b \]
The $K3$ surface can then be viewed as fiber space where the base is 
{\rm I\kern-.18em P}$^1$
and the fiber is the torus. The affine coordinate on the
{\rm I\kern-.18em P}$^1$ is         
denoted by $z$. The elliptic fibration is specified by
\[ y^2=x^3+f(z)x+g(z)\]
where $f$ is of degree 8 and $g$ is of degree 12.

If we consider $E_8\times E_8$ heterotic strings compactified on $T^2$ with 
no Wilson lines turned on so that $E_8\times E_8$ is unbroken, then it's
straightforward to show that the F-theory dual is given by the two parameter
family $(\alpha, \beta )$
\[ y^2=x^3+\alpha z^4x+(z^5+\beta z^6+z^7).\]
Moreover, in the large $\alpha$ and $\beta$ limit the complex 
structure of the torus
\( y^2=x^3+\alpha z^4x +\beta z^6 \) is the same as that of the $T^2$ upon
which the heterotic string is compactified. Given this map we can easily obtain
the F-theory encoding of the $K3$ geometry of heterotic string compactification
to 6 dimensions. We simply view $K3$ on the heterotic side  
as an elliptic fibration over {\rm I\kern-.18em P}$^1$ whose affine coordinate
we denote
by $z^{\prime}$. Now to obtain the corresponding Calabi-Yau threefold
on the F-theory side we require $\alpha$ and $\beta$ to be functions of 
of $z^{\prime}$ of degree 8 and 12 respectively, thus taking care of the 
data needed to specify the $K3$ geometry of heterotic strings in the F-theory
setup.

Further, to specify the bundle data we make the coefficients of other powers
of $z$ functions of $z^{\prime}$. Namely, Morrison and Vafa in \cite{MV2}
argued that the F-theory dual for the heterotic string configuration which
has $d_1$ zero size instantons embedded in the first $E_8$ and $d_2$ zero
size instantons in the second $E_8$ is given by
\begin{equation}
y^2=x^3+f_8(z^{\prime})z^4x+g_{d_1}(z^{\prime})z^5+g_{12}(z^{\prime})z^6
+g_{d_2}(z^{\prime})z^7 
\label{eq:duals}
\end{equation}
where the subscripts denote the degree of the polynomial.

If we wish to have no fivebranes \cite{DMW} on the heterotic side 
we need to have
$d_1+d_2=24$. We will write $d_{1, 2}=12\pm n$. For this to make sense globally
we must take $(z, z^{\prime})$ to parametrize the rational ruled surface 
{\rm I\kern-.18em F}$_{n}$. Then the deformation away from zero size 
instantons to a finite
size is written in the following form:
\begin{equation}
y^2=x^3+\sum_{k=-4}^{4}f_{8-nk}(z^{\prime})z^{4+k}x+\sum_{k=-6}^{6}
g_{12-nk}(z^{\prime})z^{6+k}
\label{eq:dual}
\end{equation}
with the condition that negative degree polynomials are set to zero. 
The moduli with $k<0$ specify the first $E_8$ bundle data,
and those with $k>0$ correspond to the second $E_8$.
 
For our purposes, it will be more convenient sometimes to use homogeneous 
coordinates for the {\rm I\kern-.18em P}$^1$'s and for the torus, representing
the latter
as a curve in a weighted projective space {\rm I\kern-.18em P}$_2^{(1, 2, 3)}$
with homogeneous coordinates $w, x, y$  and a {\bf C}$^\star$ acting on them as
$(w, x, y)\mapsto (\nu w, \nu^2 x, \nu^3 y)$. The Weierstrass equation
in homogeneous coordinates takes the form
\begin{equation}
y^2=x^3+f(s,t,u,v)xw^4+g(s,t,u,v)w^6
\label{eq:dualh}
\end{equation}
where [$s,t$] and [$u,v$] are homogeneous coordinates of the base and the
fiber {\rm I\kern-.18em P}$^1$, respectively, of the
{\rm I\kern-.18em F}$_{n}$.

Let us also recall some generalities about the spectrum of F-theory
compactifications on Calabi-Yau threefolds \cite{MV2}. Suppose we 
are given an elliptic Calabi-Yau threefold with Hodge numbers
$h_{11}(X)$ and $h_{21}(X)$ and a base $B$ which has a certain Hodge
number $h_{11}(B)$. In six dimensions we will have vector multiplets,
hypermultiplets and tensor multiplets whose number we will denote
$V$, $H$ and $n_T$, respectively.

The scalars in tensor multiplets are in one to one correspondence with the 
K\"{a}hler classes of $B$ except for the overall volume of 
$B$ which corresponds
to a hypermultiplet \cite{V:F}. Hence, we have
\begin{equation}
n_T=h_{11}(B)-1
\label{eq:nt}
\end{equation}   
Next, we use the fact that upon further compactification on $T^2$ the 
F-theory model in question becomes equivalent to the type IIA on the same
Calabi-Yau. Going to the Coulomb phase of vector multiplets in the 4D
sense we learn that
\( H^0=h_{21}(X)+1 \)
and
\begin{equation}
r(V)=h_{11}(X)-h_{11}(B)-1
\label{eq:rank}
\end{equation}
where $r(V)$ denotes the rank of the vector multiplets and $H^0$ denotes 
the number of neutral hypermultiplets.
There is also an anomaly cancellation condition which requires that
\cite{Scw:an}
\begin{equation}
H-V=273-29n_T
\label{eq:anc}
\end{equation}
where $H$ and $V$ denote the total number of hypermultiplets and 
vector multiplets respectively.

Note that the base of the elliptic fibration need not be one of
the {\rm I\kern-.18em F}$_{n}$'s In fact, that would limit us to having only
one 
tensor multiplet in 6 dimensions
(since $h_{11}$({\rm I\kern-.18em F}$_{n})=2$).
Actually we know \cite{MV2} that the requirement of having $N=1$
supersymmetry in six dimensions limit us  to a base which is either
an Enriques surface, a blowup of {\rm I\kern-.18em P}$^2$ or 
{\rm I\kern-.18em F}$_{n}$, or a surface
with orbifold singularities whose resolution is one of the above.
In our construction surfaces which are blowups of {\rm I\kern-.18em F}$_{n}$'s
will be of importance.

\pagebreak

\section{Nonperturbative Vacua and Reflexive \\ Polyhedra}

\subsection{ Blowing up the base}

Since our task now is to obtain models with more than one tensor 
multiplet in six dimensions, we need to blow up the base of the elliptic
Calabi-Yau in accordance with~(\ref{eq:nt}) in Section 2.

The starting point is one of the {\rm I\kern-.18em F}$_{n}$ models reviewed in
Section 2.
It is described by the following table:
\begin{equation}
\begin{array}{lllllccl}    
         & s & t & u & v & x & y & w \\
\lambda\ & 1 & 1 & n & 0 & 2(n+2) & 3(n+2) & 0 \\
\mu\     & 0 & 0 & 1 & 1 & 4      & 6      & 0 \\
\nu\     & 0 & 0 & 0 & 0 & 2      & 3      & 1 \\
\end{array}   
\label{m:n}
\end{equation}
which specifies the exponents with which $(\lambda ,\mu ,\nu)$
act on the homogeneous coordinates $s, t, u, v, x, y, w$.  
The Calabi-Yau threefold is obtained by starting with these 
homogeneous coordinates, removing the loci $\{s=t=0\}, \{u=v=0\},
\{x=y=w=0\}  $, taking the quotent by $({\bf C}^\star)^3$,
and restricting to the solution set of~(\ref{eq:dualh}) . 
\subsubsection{$n=0$}

Let's consider the case $n=0$ which corresponds to symmetric embedding
of 24 instantons in the two $E_8$'s. Our Calabi-Yau threefold in this case
has $(h_{11}, h_{21})=(3, 243)$. We want to blow up one point in the base
{\rm I\kern-.18em F}$_0$ in order to obtain a model with $h_{11}(B)=2+1=3$ and
$n_T=2$ according to~(\ref{eq:nt}).

To do that we introduce a new coordinate $p$ along with a {\bf C}$^\star$
which acts on the coordinates as
\begin{equation}
(p, s, t, u, v, x, y, w)\mapsto(\sigma p, \sigma s, t, \sigma u, v,
\sigma^6 x, \sigma^9 y, w)
\label{1bu}
\end{equation}
We claim that this model is dual to the heterotic one with 12 instantons 
in the first E$_8$ and 11 in the second E$_8$. Indeed, the equation 
for the Calabi-
Yau threefold in this space looks like the following (Recall that 
$z^\prime$ and
$z$ are affine coordinates of the base and fibre  {\rm I\kern-.18em P}$^1$'s
of the 
{\rm I\kern-.18em F}$_{n}$ respectively, while $[s,t]$ and $[u,v]$ are the
corresponding 
homogeneous coordinates.):
\begin{equation}
y^2=x^3+\sum_{k=-4}^{4}\sum_{l=0}^{8}f_{kl}z^{\prime l}z^{4+k}p^{8-l-k}x
+\sum_{k=-6}^{6}\sum_{l=0}^{12}g_{kl}z^{\prime l}z^{6+k}p^{12-l-k}
\label{eq:n=0}
\end{equation}
We see that certain terms of the original equation are necessarily 
suppressed by the requirement that we obtain a Calabi-Yau hypersurface here.
In particular the highest power of $z^{\prime}$ multiplying $z^{5}$
gives us the number of instantons in the first $E_8$, $d_1$ and that 
multiplying
$z^7$ supplies us with $d_2$. Hence, our model should correspond to
$(d_1, d_2)=(12, 11)$, which is in agreement with having $n_T=2$. We also
expect $h_{11}$ of our threefold to go up by 1, and it indeed does,
yielding $(h_{11}, h_{21})=(4, 214)$ which is in perfect agreement
with the anomaly cancellation condition~(\ref{eq:anc}).

Also note that according to~(\ref{eq:rank}) 
\begin{displaymath}
r(V)=4-3-1=0,
\end{displaymath}
and there is no gauge group in six dimensions: only extra tensor multiplet
appears in the spectrum.
We can continue this process and try to get rid of one more instanton.
That can be done by replacing $\sigma u$ in the right-hand side 
of~(\ref{1bu}) by $\sigma^2 u$, or by considering the following model:
\begin{displaymath}
\begin{array}{llllllccl}
         & p & s & t & u & v & x & y & w \\
\lambda\ & 0 & 1 & 1 & 0 & 0 & 4 & 6 & 0 \\
\mu\     & 0 & 0 & 0 & 1 & 1 & 4 & 6 & 0 \\
\nu\     & 0 & 0 & 0 & 0 & 0 & 2 & 3 & 1 \\
\sigma\  & 1 & 1 & 0 & 2 & 0 & 8 & 12 & 0 \\
\end{array}
\end{displaymath} 
Indeed, in this case the equation for our threefold takes the form:
\begin{equation}
y^2=x^3+\sum_{k=-4}^{4}\sum_{l=0}^{8}f_{kl}z^{\prime l}z^{4+k}p^{8-l-2k}x
+\sum_{k=-6}^{6}\sum_{l=0}^{12}g_{kl}z^{\prime l}z^{6+k}p^{12-l-2k}
\label{eq:2bu}
\end{equation}
Again we immediately notice that the terms must be further suppressed by
in order to obtain a Calabi-Yau hypersurface. In particular, the maximal power
of $z^{\prime}$ multiplying $z^7$ is now 10 which tells us that now we have
only 10 instantons in the second $E_8$ on the heterotic side. Ten instantons
are enough to break $E_8$ completely, so we don't expect any gauge group
in six dimensions. ~(\ref{eq:rank}) gives us:
 \[ r(V)=5-4-1=0, \]
since the Calabi-Yau now has $(h_{11}, h_{21})=(5,195)$.
 
We can continue this process in an obvious way, namely to obtain the model
dual to the heterotic vacuum with $(d_1, d_2)=(12, 12-q)$ and $n_T=q+1$
we consider the set of weights:
\begin{displaymath}
\begin{array}{llllllccl}
         & p & s & t & u & v & x & y & w \\
\lambda\ & 0 & 1 & 1 & 0 & 0 & 4 & 6 & 0 \\
\mu\     & 0 & 0 & 0 & 1 & 1 & 4 & 6 & 0 \\
\nu\     & 0 & 0 & 0 & 0 & 0 & 2 & 3 & 1 \\
\sigma\  & 1 & 1 & 0 & q & 0 & 2(q+2) & 3(q+2) & 0 \\
\end{array}
\end{displaymath}
and define the Calabi-Yau exactly as before.
The equation now is 
\begin{equation}
y^2=x^3+\sum_{k=-4}^{4}\sum_{l=0}^{8}f_{kl}z^{\prime l}z^{4+k}p^{8-l-qk}x
+\sum_{k=-6}^{6}\sum_{l=0}^{12}g_{kl}z^{\prime l}z^{6+k}p^{12-l-qk}
\label{eq:qbu}
\end{equation}
with the understanding that terms containing negative powers are 
absent.

We see that requiring that all powers of $p$ are nonnegative amounts  
to the fact that the maximal power of $z^{\prime}$ multiplying $z^7$ 
in the last
sum in the equation is only $12-q$ (the maximal power of $z^{\prime}$
multiplying $z^5$ is still 12) which we interpret as having 12 instantons
in the first $E_8$ and $12-q$ instantons in the second $E_8$ in the heterotic
dual. 
 
In the same way, we can build duals to vacua with different instanton numbers
in the first $E_8$ also. We simply introduce one more coordinate and one more
{\bf C}$^\star$ which acts on it. Namely, we consider the model 
defined by the set of
weights
\begin{equation}
\begin{array}{lllllllccl}
         & r  & p & s & t & u & v & x & y & w \\
\lambda\ & 0  & 0 & 1 & 1 & 0 & 0 & 4 & 6 & 0 \\
\mu\     & 0  & 0 & 0 & 0 & 1 & 1 & 4 & 6 & 0 \\
\nu\     & 0  & 0 & 0 & 0 & 0 & 0 & 2 & 3 & 1 \\
\sigma\  & 0  & 1 & 1 & 0 & q & 0 & 2(q+2) & 3(q+2) & 0 \\
\xi\     & 1  & 0 & 1 & 0 & 0 & q^{\prime}& 2(q^{\prime}+2) & 3(q^{\prime}+2)
 & 0 \\
\end{array}
\label{m:qq`bu}
\end{equation} 
and restrict our set to the solutions of
\begin{eqnarray}
\lefteqn{y^2=x^3+\sum_{k=-4}^{4}\sum_{l=0}^{8}f_{kl}
z^{\prime l}z^{4+k}p^{8-l-qk}
r^{8-l+q^{\prime}k}x} \nonumber  \\ 
& & +\sum_{k=-6}^{6}\sum_{l=0}^{12}g_{kl}z^{\prime l}z^{6+k}p^{12-l-qk}
r^{12-l+q^{\prime}k}
\label{eq:qq`bu}
\end{eqnarray}
Again all terms containing negative powers are understood to be nonexistent.
In particular having only nonnegative powers of $p$ and $r$ requires
that the maximal power of $z^{\prime}$ multiplying $z^5$ be $12-q^{\prime}$ 
and that
for $z^7$ be $12-q$. This, according to our interpretation, shows us
that on the heterotic side we have $(d_1, d_2)=(12-q^{\prime}, 12-q)$.

\subsubsection{$n\neq 0$}

Now suppose we had $12+n$ instantons in the first $E_8$ and $12-n$ instantons
in the second $E_8$ in our perturbative heterotic vacuum. The F-theory
model dual to it was described by~(\ref{m:n}) and~(\ref{eq:dual}). As before,
we want to blow up the base {\rm I\kern-.18em F}$_{n}$ in order to produce 
duals to the vacua with more than one tensor multiplets in six dimensions. 
The way we do it in this case 
is very similar to how it was done in the previous subsection. Namely,
we claim that the model
\begin{equation}
\begin{array}{llllllccl}
           & p & s & t & u & v & x      & y      & w \\
\lambda\   & 0 & 1 & 1 & n & 0 & 2(n+2) & 3(n+2) & 0 \\
\mu\       & 0 & 0 & 0 & 1 & 1 & 4      &   6    & 0 \\
\nu\       & 0 & 0 & 0 & 0 & 0 & 2      &   3    & 1 \\
\sigma\    & 1 & 1 & 0 & q & 0 & 2(q+2) & 3(q+2) & 0 \\
\end{array}
\label{m:nlq}
\end{equation}
is dual to $(d_1, d_2)=(12+q, 12-n)$ heterotic vacuum if $q<n$ and
to $(d_1, d_2)=(12+n, 12-q)$ if $q>n$. Indeed, the equation in this case 
has the following form:
\begin{equation}
y^2=x^3+\sum_{k=-4}^{4}\sum_{l=0}^{8-nk}f_{kl}z^{\prime l}z^{4+k}p^{8-l-qk}x
+\sum_{k=-6}^{6}\sum_{l=0}^{12-nk}g_{kl}z^{\prime l}z^{6+k}p^{12-l-qk}
\label{eq:nlq}
\end{equation}
So now the maximal power of $z^{\prime}$ multiplying $z^5$ is
$l_{\rm max}={\rm min}(12+n, 12+q)$ and the maximal power multiplying
$z^7$ is $l_{\rm max}={\rm min}(12-n, 12-q)$, which amounts to our
 claim.   
 
Similarly, the set of weights
\begin{equation}
\begin{array}{llllllccl}
           & p & s & t & u & v & x      & y      & w \\
\lambda\   & 0 & 1 & 1 & n & 0 & 2(n+2) & 3(n+2) & 0 \\
\mu\       & 0 & 0 & 0 & 1 & 1 & 4      &   6    & 0 \\
\nu\       & 0 & 0 & 0 & 0 & 0 & 2      &   3    & 1 \\
\sigma\    & 1 & 1 & 0 & 0 & q^{\prime} & 2(q^{\prime}+2) 
& 3(q^{\prime}+2) & 0 \\
\end{array}
\label{m:nlq`}
\end{equation}
with the equation
\begin{equation}
y^2=x^3+\sum_{k=-4}^{4}\sum_{l=0}^{8-nk}f_{kl}z^{\prime l}z^{4+k}
p^{8-l+q^{\prime}k}x
+\sum_{k=-6}^{6}\sum_{l=0}^{12-nk}g_{kl}z^{\prime l}z^{6+k} 
p^{12-l+q^{\prime}k}
\label{eq:nlq`}
\end{equation}
define the model dual to a nonperturbative heterotic vacuum with
$(d_1, d_2)=(12-q^{\prime}, 12-n)$.

We are now in a position to put the pieces together and construct a
Calabi-Yau threefold which yields F-theory/Type II model dual
to a nonperturbative heterotic vacuum with $d_1$ instantons embedded
in the first $E_8$ and $d_2$ instantons in the second $E_8$
for any $d_1+d_2<24$.

If $d_1>12$ (we can always assume that $d_2<12$) we just take
~(\ref{m:nlq}) and~(\ref{eq:nlq}) with either $n=d_1-12$ and $q=12-d_2$
or vice versa ($q=d_1-12$, $n=12-d_2$).

If $d_1<12$ we take~(\ref{m:nlq`}) and~(\ref{eq:nlq`}) with
$n=12-d_2$ and $q^{\prime}=12-d_1$. Note that we could also achieve
this result by starting with $n=0$ and applying~(\ref{m:qq`bu})
and~(\ref{eq:qq`bu}) with $q=12-d_2$ and $q^{\prime}=12-d_1$.  

\subsection{Reflexive Polyhedra and Hodge Numbers}

Here  we will analyze the manifolds constructed in the previous section.
Our main tool is Batyrev's toric approach. To a Calabi-Yau manifold
defined as a hypersurface in a weighted projective space one can
associate its Newton polyhedron, which we denote by $\Delta$. If it
happens to be reflexive, which it often (perhaps always) does,
we may define the dual or polar polyhedron which we denote by $\nabla$.
By means of a computer program we have computed the dual polyhedra
and Hodge numbers for all the examples from the previous section. 
The polyhedra exhibit a nice regular structure similar to that 
discussed in \cite{poly:CF}.

To illustrate the aforementioned structure, let's have a look at the 
first three cases. Namely, consider the dual polyhedra for the models
dual to the heterotic vacua with $(d_1, d_2)$ equal to 
$(12, 12)$, $(12,11)$ and
$(12, 10)$. The points of dual polyhedra are displayed in Table 3.1.

\pagebreak

\begin{tabular}{|c|c|c|}   \hline
 \hspace{10mm}(12,12)\hspace{10mm}   &\hspace{10mm}(12,11)\hspace{10mm}
&\hspace{10mm}(12,10)\hspace{10mm} \\ \hline\hline
  (-1, 0, 2,-1)   &  (-1, 0, 2,-1)   & (-1, 0, 2,-1)\\ 
  ( 0,-1, 1,-1)   &  ( 0,-1, 1,-1)   & ( 0,-1, 1,-1)\\ 
  ( 0, 0,-1,-1)   &  ( 0, 0,-1,-1)   & ( 0, 0,-1,-1)\\    
  ( 0, 0, 0,-1)   &  ( 0, 0, 0,-1)   & ( 0, 0, 0,-1)\\
  ( 0, 0, 0, 0)   &  ( 0, 0, 0, 0)   & ( 0, 0, 0, 0)\\
  ( 0, 0, 1,-1)   &  ( 0, 0, 1,-1)   & ( 0, 0, 1,-1)\\
  ( 0, 0, 1, 0)   &  ( 0, 0, 1, 0)   & ( 0, 0, 1, 0)\\ 
  ( 0, 0, 1, 1)   &  ( 0, 0, 1, 1)   & ( 0, 0, 1, 1)\\
  ( 0, 0, 1, 2)   &  ( 0, 0, 1, 2)   & ( 0, 0, 1, 2)\\
  ( 0, 1, 1,-1)   &  ( 0, 1, 1,-1)   & ( 0, 1, 1,-1)\\
  ( 1, 0, 0,-1)   &  ( 1, 0, 0,-1)   & ( 1, 0, 0,-1)\\  
                  &  ( 1,-1, 0,-1)   & ( 1, 1, 0,-1)\\
                  &                  & ( 1, 2, 0,-1)\\ \hline
\end{tabular}
\bigskip

{\bf Table 3.1:} The dual polyhedra for given $(d_1, d_2)$ on the heterotic    
side.
\bigskip

The following observations summarize the structure of the polyhedra:
\begin{enumerate}
 \item{For each polyhedron the points from the second up to tenth (counting
from above) lie in the hyperplane $x_1=0$ and are the same. These points
themselves form a reflexive polyhedron which is dual of the polyhedron
for the $K3$ of the original fibration.}
 \item{The points from the third up to ninth in each case form a two-
dimensional reflexive polyhedron, $^2\nabla$, which is a triangle. 
This  $^2\nabla$ is the dual polyhedron of the torus 
{\rm I\kern-.18em P}$_2^{(1,2,3)}$
of the elliptic fibration.}
 \item{The first eleven points in all three cases are exactly the same.
The second polyhedron differs from the first by the addition of the 
point (1,-1, 0,-1) which lies `above' the hyperplane of the $K3$.
The third polyhedron differs from the first by the addition of two points
`above' the $K3$.}
\end{enumerate} 
 We have checked that this pattern persists in all other examples.\footnote[1]
{In the cases where both $d_1$ and $d_2$ are less than 10, and so there are
unbroken subgroups in both $E_8$'s, the subpolyhedron defined by $x_1=0$
is still the dual of a $K3$ manifold, but the way the elliptic fibration
structure is visible inside it is trickier. Namely, the two-dimensional
subpolyhedron which here lies in the plane $x_1=0, x_2=0$ and which is 
a dual polyhedron of a torus gets tilted in some cases and is no longer 
visible in other cases.} 
Namely,
when we blow up one more point in the base of the elliptic fibration
thus decreasing the total number of instantons on the heterotic side
by one, one more point appears `above' the $K3$ hyperplane. These points
always line up, that is they have the form\footnote[2]{Note that these points
are all evenly spaced and lie in a straight line. It is easily verified that 
they actually lie on an edge of the dual polyhedron. Following \cite{KMP},it 
may be argued that such a situation signals the appearance of an enhanced 
gauge symmetry. However, this argument applies only when the 
genus of the corresponding 
curve of singularities is greater than 1. In our case, the genus of this curve 
is exactly one, as can be seen by going to the original polyhedron and 
counting the number of integral points interior to the 2-face dual to this
edge.}    
\[ (1, k, 0,-1),\; k=0, 1,\ldots , n_T-1\]
with a few exceptions  (e. g. in $(12, 11)$ case
we have points $(1,-1, 0,-1)$ and $(1, 0, 0,-1)$). This is in precise
agreement (up to a change of basis) with the observation of \cite{poly:CF}.

 Also as $d_1$ or $d_2$ become less than 10 the three-dimensional polyhedron
which we associate to the $K3$ of the fibration begins to acquire additional
points signaling the appearance of unbroken gauge groups 
which have perturbative
interpretation on the heterotic $E_8\times E_8$ side. This is again in 
agreement with what we would expect from the heterotic point of view. Indeed,
9 instantons and less is not enough to break $E_8$ completely, 
and some subgroup
of either (or both) of the $E_8$'s should appear in the perturbative part
of the spectrum, which we know is visible in the generic $K3$ fibre of 
the $K3$ fibration on the Type II side \cite{Asp}. Moreover, the group
which we observe in the $K3$ is precisely the commutant of the  maximal
subgroup of $E_8$ which can be completely broken by the given number of
instantons, i.e. we are dealing with maximally Higgsed case. 

It is interesting to note that projection of the points in Table 3.1  
onto the first two coordinates yields the fan of the base of the elliptic 
fibration - thus in the case of $(12,12)$ instantons in the two $E_8$'s, 
we get the points $(1,0),(0,1),(0,0),(-1,0),(0,-1)$, which we recognise as 
the fan of {\rm I\kern-.18em F}$_{0}$. Creating a tensor multiplet by removing
an instanton, we find that the base acquires an additional point, namely
$(1,-1)$. To see that this corresponds to a blowup of
{\rm I\kern-.18em F}$_{0}$, we use a standard result in toric
varieties\cite{Fulton} - insertion of the sum of two adjacent vectors in the 
fan of a nonsingular toric surface yeilds a blowup of the surface. In our
case, we simply note that $(1,-1)=(1,0)+(0,-1)$, so that we do indeed have 
{\rm I\kern-.18em F}$_{0}$ blown up at one point. We find this pattern to hold
in all the cases we have studied. 
Furthermore, we can compute
$h_{11}$ of the base, which we know from \cite{MV2} to be $n_T + 1$. Once again
we find exact correspondence between geometry and physics.
For {\rm I\kern-.18em F}$_{n}$ and blowups thereof, $h_{11}=b_2$. 
The formula for
$b_{2}$ of a toric surface is $b_2 = d_1 - 2d_0$, where $d_i$ is the number of
$i$ dimensional cones. Thus for {\rm I\kern-.18em F}$_{0}$, the point 
$(0,0)$ generates the only zero dimensional cone in the fan while all the
other points generate one dimensional cones, so we find that 
$b_2$({\rm I\kern-.18em F}$_{0}$)$= 4 - 2 = 2$, which is one more than the
number of tensor multiplets in the $(12,12)$ model. Also, blowing up the base
increases the number of one dimensional cones by one while leaving the number
of zero dimensional cones unchanged, so that $b_2$ of the base increases by one
each time.
     
Having thus outlined the general picture let's look at some of the
examples.
\begin{enumerate} 
\item $(d_1, d_2)=(12, 12)$.
$(h_{11}, h_{21})=(3, 243)$
As we've already seen, there is only one point in the dual polyhedron
above the $K3$ hyperplane, which corresponds to $n_T=1$ as becomes to 
a perturbative vacuum. $h_{11}(B)=n_T+1=2$, 
$r(V)=h_{11}(X)-h_{11}(B)-1=0$, so there is no gauge group in 6D.
Indeed, the three-dimensional dual polyhedron corresponding to the
generic fiber of the fibration shows no sign of a gauge group
according to the results of \cite{poly:CF}. 

If we count the moduli
specifying the gauge bundle data, we obtain 114 for each $E_8$.
At this point, though, we still have the freedom of one rescaling for $z$ 
\cite{Ber}. Also, one of the moduli corresponds to a deformation of 
{\rm I\kern-.18em F}$_{0}$ to {\rm I\kern-.18em F}$_{2}$. 
Thus we have 112 degrees of freedom, which is 
exactly the quaternionic dimension of the moduli space of 12
$E_8$ instantons.
\item $(d_1, d_2)=(12,11)$.
$(h_{11}, h_{21})=(4, 214)$
One point is added `above' the $K3$, signalling the appearance of
one additional tensor multiplet. $h_{11}(B)=3$, $r(V)=4-3-1=0$.
The $K3$ polyhedron hasn't changed, so we don't expect any perturbative
heterotic contributions to the gauge group. On the other hand, one 
less instanton results in a tensor multiplet in 6D, which becomes a 
vector multiplet in 4D. So we expect no nonperturbative 
heterotic contribution to
the 6D gauge group, which is in agreement with what we find for $r(V)$.
Also $H=H^0=215$, $V=0$, and the anomaly cancellation condition~(\ref{eq:anc})
is satisfied. 

The number of moduli corresponding to the first $E_8$ is, 
of course, the same as in the previous example. 
However, the number of moduli
specifying the second $E_8$ bundle has dropped to 83 due 
to the suppression of certain terms which would yield negative powers
of $p$ in~(\ref{eq:qbu}). Taking into account the $z$ scaling
freedom makes it 82, which is the dimension of the moduli space of 
11 $E_8$ instantons.  
\item $(d_1, d_2)=(12,10)$.
$(h_{11}, h_{21})=(5,185)$. 
$K3$ is still the same. We see three points 
`above' which implies $n_T=3$, $h_{11}(B)=4$. $r(V)=5-4-1=0$,  
$H=H^0=186$, $V=0$, anomaly cancellation obviously holds.
We now obtain 54 moduli specifying the bundle data in the second $E_8$. 
However, one of the moduli corresponds to a deformation of 
{\rm I\kern-.18em F}$_{2}$ to {\rm I\kern-.18em F}$_{0}$.
There is also one scaling degree of freedom for $z$, so that the number
of degrees of freedom is 52, which is exactly the dimension of the moduli
space of 10 $E_8$ instantons.  
\item $(d_1, d_2)=(12,9)$.
$(h_{11}, h_{21})=(8,164)$. 
4 points correspond to $n_T=4$, $h_{11}(B)=5$, and hence
$r(V)=8-5-1=2$. Thus, we expect a rank 2 perturbative (on the heterotic side)
gauge group. Indeed, 9 instantons break an $E_6$ subgroup completely
leaving us with an unbroken $SU(3)$, which can also be read off from the $K3$
dual polyhedron. Also, the $SU(3)$ comes with no charged matter in this case,
so $H=H^0=165$, $V={\rm dim}(SU(3))=8$, and the anomaly cancellation reads
$165-8=273-29\cdot 4$.

Let's also see what the moduli counting tells us. $k>0$ 
terms in ~(\ref{eq:qbu})
with $q=3$ now number 30 (where the rescaling freedom has been taken into 
account). 9 instantons on the heterotic side fit into an $E_6$, which
has dual Coxeter number 12 and dimension 78. So, the moduli space of 9
instantons embedded in the $E_6$ has quaternionic dimension 
$hd_2-{\rm dim}(G)=12\cdot 9-78=30$ in complete agreement with our number 
of complex moduli on F-theory side. 
\item $(d_1, d_2)=(12,8)$.
$(h_{11}, h_{21})=(11,155)$. 
One more point adds up above the $K3$ hyperplane. Points are also added
to the $K3$ itself. Using the results of \cite{poly:CF}, we see an unbroken
$SO(8)$ appearing. Also, we have $n_T=5$, $h_{11}(B)=6$, $r(V)=11-6-1=4$,
in agreement with the rank of $SO(8)$. Again, we get no charged matter.
Hence, $H=H^0=156$, $V={\rm dim}(SO(8))=28$, and
$156-28=273-5\cdot 29$ as required by the anomaly cancellation.   

The equation~(\ref{eq:qbu}) now has 20 moduli with $k>0$. 
On the other hand, 8 instantons
can break an $SO(8)$ subgroup of the second $E_8$ completely. $SO(8)$
has dual Coxeter number 6 and dimension 28. So, for the dimension of
the moduli space of 8 $SO(8)$ instantons we have $6\cdot 8-28=20$, again
precisely matching the number of F-theory moduli. 
\item $(d_1, d_2)=(12,7)$.
$(h_{11}, h_{21})=(12,150)$. 
In the same way, we get $n_T=6$, $h_{11}(B)=7$. The shape of the 
three-dimensional polyhedron corresponding to $K3$ is exactly that
identified with the presence of an unbroken $E_6$ in \cite{poly:CF}.
We have to be more careful, though. In \cite{AG:32}, \cite{Ber} it was 
shown that the 
$E_6$ singularity can also correspond to an unbroken $F_4$. We know
also that on the heterotic side  
breaking can proceed up to $F_4$ \cite{Kap:pr}. On the other hand, we 
obtain $r(V)=12-7-1=4$, which is in agreement with having an unbroken $F_4$
in 6D. In addition, the anomaly cancellation condition gives us (supposing
$H=H^0$): $151-52=273-29\cdot 7$ (${\rm dim}( F_4)=52$), which
is in precise agreement with having a matter-free $F_4$. 

Counting the number of $k>0$ moduli in~(\ref{eq:qbu}) with $q=5$ now
yields 14. 7 instantons fit in a $G_2$ subgroup of the $E_8$. $G_2$
has dual Coxeter number 4 and dimension 14 giving us $4\cdot 7-14=14$
for the moduli space of 7 instantons on the heterotic side.
\item $(d_1, d_2)=(12,6)$.
$(h_{11}, h_{21})=(15,147)$. 
One more point above $K3$ appears. All other points are exactly as
in $(12, 7)$ case. That is, the subpolyhedron corresponding to the 
$K3$ still shows us the presence of an $E_6$ singularity in the Higgs
branch. This time, though, we know from \cite{MV2}, \cite{AG:32}, 
\cite{Ber} and from 
the argument on the heterotic side that an $E_6$ gauge group actually
makes its appearance. Indeed, we have $n_T=7$, $h_{11}(B)=8$, and,
hence, $r(V)=15-8-1=6$ and, moreover, anomalies cancel
precisely when $H=H^0=148$ and $V={\rm dim}(E_6)=78$. 

Here we used~(\ref{eq:qbu}) with $q=6$ which has 10 $k>0$ complex
moduli. 6 instantons, on the other hand, fit into an $SU(3)$ subgroup
which has dual Coxeter number 3 and dimension 8 yielding thus
$3\cdot 6-8=10$ quaternionic moduli on the heterotic side and showing
precise agreement.
\item $(d_1, d_2)=(12,5)$.
$(h_{11}, h_{21})=(17,145)$. 
We expect to see an unbroken $E_7$ with half a 56 hypermultiplet \cite{MV2}.
Our polyhedron acquires one more point above the $K3$ hyperplane, 
as well as a certain
number of points in the $K3$ subpolyhedron itself, which we can see to 
signal the appearance of $E_7$ \cite{poly:CF}. We have $n_T=8$, $h_{11}(B)
=9$ and $r(V)=17-9-1=7$. Moreover, the anomaly cancellation condition
works precisely when $H=H^0+H^c=146+{1\over 2}\cdot 56 =174$ and
$V={\rm dim}(E_7)=133$, which is perfectly consistent with what
we expected.  

As to the number of moduli, we get 7 of them from our equation
(counting only those with $k>0$) which matches exactly what we obtain on
the heterotic side for 5 instantons: they fit in $SU(2)$, which has dual
Coxeter number 2 and dimension 3, producing
$2\cdot 5-3=7$ quaternionic moduli.
B
\item $(d_1, d_2)=(12,4)$.
$(h_{11}, h_{21})=(18,144)$. 
In comparison to the previous case, the polyhedron gets one more point
above the $K3$. All the rest carry over. So now $n_T=9$, $h_{11}(B)=10$.
We expect $E_7$ without any charged matter. Indeed, the $K3$ subpolyhedron 
is the same as in $(12, 5)$ case and corresponds to $E_7$,  and for 
the rank of the gauge group we obtain $r(V)=18-10-1=7$. Again the 6D anomalies
cancel exactly when $H=H^0=145$ and $V={\rm dim}(E_7)=133$.

The equation~(\ref{eq:qbu}) with $q=8$ yields 5 moduli. 
4 instantons fit in an $SU(2)$ subgroup of the second $E_8$.  So 
we obtain $2\cdot 4-3=5$ moduli on the heterotic side also.  
\item $(d_1, d_2)=(12,0)$.
$(h_{11}, h_{21})=(23,143)$. 
Now there are 13 points above the $K3$ subpolyhedron meaning that
$n_T=13$ and $h_{11}(B)=14$. The $K3$ subpolyhedron itself acquires 
some additional points also exhibiting the structure which we claim
corresponds to $E_8$. For the rank of the gauge group we obtain
$r(V)=23-14-1=8$, and the anomalies will cancel if $H=H^0=144$ and
$V={\rm dim}(E_8)=248$, which is in accord with having matter-free
$E_8$. This again agrees perfectly with our expectations. 

The counting of moduli now is trivial and gives 0 on both sides.
\end{enumerate}
\medskip

Using the blowup procedure described in the previous subsection, 
we can also construct vacua dual to heterotic ones with $d_1<12$ as well
as with $d_1>12$. In all of these cases, 
the rank of the total gauge group obtained from~(\ref{eq:rank})
is in precise agreement with what can be deduced from the heterotic side
(assuming maximal possible Higgsing).
Also, the Hodge numbers work in such a way that the anomaly cancellation
condition is always satisfied. Moduli counting shows perfect
agreement as well. This can be regarded as a strong check of the     
proposed algorithm of constructing F-theory duals to nonperturbative heterotic
vacua.

\subsection{Unhiggsing $E_8$}
While describing various examples in the previous subsection, we left
aside the cases when either $d_1$ or $d_2$ (or both) become less than 4.
We know that in this situation, the instantons cannot be of finite size
(various $D$-terms in six dimensions wouldn't let us inflate them)  
\cite{W:si}. This corresponds to the fact that the unbroken gauge symmetry is
$E_8$. Let's see what happens if we try to consider models with 
$d_2<4$.
\begin{enumerate}
\item $(d_1, d_2)=(12,3)$.
In this case our threefold has $(h_{11}, h_{21})=(23, 143)$
exactly as in $(12,0)$ case. The manifold itself is apparently 
different, though: there are 3 fewer points above the three-dimensional
subpolyhedron which we associate to $K3$, and those points are not being
erased from a codimension one face. The rest of the points stay exactly
where they were leaving the three-dimesional subpolyhedron unchanged.
We are thus led to the conclusion that unbroken $E_8$ is present (indeed,
zero size instantons can't break it) and also that there appear to be only 10
tensor multiplets in 6 dimensions (again, one tensor multiplet corresponding 
to one point with $x_1=1$ ). For the rank of the total
gauge group we obtain: $r(V)=8$, and $h_{11}(B)=23-8-2=13$ strongly
suggesting that there are actually 13 tensor multiplets - so that there are
3 extra tensors which are somehow not visible as points in the polyhedron.     

It's interesting also to look at the moduli on the F-theory side and
see that their number is consistent with having 3 instantons
in the second $E_8$. Indeed, in~(\ref{eq:qbu}) with $q=9$ there are only
4 nonzero terms with $k>0$ (they are $g_{1l}$  for $l=0,1,2,3$), 1 of which
is irrelevant. This gives us 3 moduli corresponding to the positions
of 3 (point-like) instantons on the $K3$ on the heterotic side.   

\item $(d_1, d_2)=(12,2)$.
Again we see $(h_{11}, h_{21})=(23, 143)$ The polyhedron is exactly as before 
except for one point with $x_1=1$ which joins the 10 present in $(12, 3)$
case making their number 11. Thus it appears that $n_T=11$, $h_{11}(B)=12$
and $r(V)=8$, yielding a discrepancy of 2 tensor multiplets.

Exactly as in the previous example, we have 2 moduli on the F-theory side
corresponding to the positions of 2 (small) instantons. 
\item $(d_1, d_2)=(12,1)$.
Same story:$(h_{11}, h_{21})=(23, 143)$, $n_T=12$, $h_{11}(B)=13$ and
$r(V)=8$, so that we are missing one tensor multiplet.
\end{enumerate}
\medskip

In the above examples, the instantons happen to be small because they 
are not allowed to be of finite size. Following \cite{AG:32}, we know that
we can't have smooth Calabi-Yau's corresponding to small instantons. In the 
case of small $SO(32)$ instantons, one can smooth out the singularity without
blowing up the base, but for small $E_8$ instantons, smoothing out the
singularity necessarily involves blowing up the base - thus the small
instantons become tensor multiplets. 
   
We were able to construct an algorithm similar to that of the previous
section which allows us to build duals to heterotic vacua with enhanced $E_8$
gauge symmetry\footnote{This procedure naturally
shrinks all the instantons in the given $E_8$.}. 
It works as follows.

Given a manifold constructed as described before, add one 
more new coordinate
along with an additional {\bf C}$^\star$ which acts on the coordinates as:
\begin{equation}
(a, u, x, y)\mapsto (\rho a, \rho^6 u, \rho^{14} x, \rho^{21} y)
\label{sc:si}
\end{equation} 
 the action on other coordinates being trivial.
Then remove suitable loci and restrict to the set of solutions of
\begin{equation}
y^2=x^3+\sum_{k=-4}^{4}\sum_{l=0}^{12-nk}f_{kl}z^{\prime l}
z^{4+k}\ldots a^{4-6k}
x+\sum_{k=-6}^{6}\sum_{l=0}^{12-nk}g_{kl}z^{\prime l}z^{6+k}\ldots a^{6-6k}
\label{eq:si}
\end{equation} 
where $\ldots$ stand for any other coordinates which can be present.

We claim that this represents the dual to a heterotic vacuum where the second
$E_8$ is unhiggsed. Indeed, it's easy to see 
that the requirement that only nonnegative powers of the new
coordinate $a$ are present leads to the absence of all powers 
of $z$ bigger than
4 in the second term  and all powers of $z$ bigger than 7 in the third term
of RHS of~(\ref{eq:si}), 
so that by Tate's algorithm\cite{MV2}, the second $E_8$ is unhiggsed.
Also, if we count F-theory moduli, we'll find precisely the number
equal to the number of instantons in the second $E_8$, and they are 
obviously interpreted as corresponding to the position of point-like 
instantons.

The general picture which emerges is best illustrated by the following 
examples.
\smallskip
 
Consider a dual to the heterotic model with 12 instantons in the first
$E_8$ and no instantons in the second $E_8$. We know that the Calabi-Yau
in this case has $(h_{11}, h_{21})=(23, 143)$ and its dual polyhedron
consists of: 
\begin{itemize}
\item A set of points with $x_1=0$ which happens to be a dual 
polyhedron of the $K3$ exibiting unbroken $E_8$ symmetry 
\item One point with $x_1=-1$
\item A set of points $(1, 0, 0,-1), (1, 1, 0,-1),\ldots (1, 12, 0,-1)$
\end{itemize}

Now let's look at the dual to a heterotic model with 12 finite size instantons
in the first $E_8$ and 12 small instantons in the second $E_8$.
The Calabi-Yau threefold turns out to have exactly the same Hodge numbers:
$(h_{11}, h_{21})=(23, 143)$ and its dual polyhedron consists of:
\begin{itemize}
\item Same set of points as before with $x_1=0$
\item Same point with $x_1=-1$
\item Only one point with $x_1=1$: $(1, 0, 0,-1)$
\end{itemize}

This picture holds in general: take a $(d_1, 0)$ vacuum. It has an unbroken
$G\times E_8$ gauge group where $G$ corresponds to the maximally
Higgsed case in the first $E_8$. Compare it to any $(d_1, d_2)$
vacuum where all $d_2$ instantons are point-like. The corresponding 
Calabi-Yau threefold is characterized by the same Hodge numbers, and
the dual polyhedron differs by the absence of the points
$(1, 24-(d_1+d_2)+1, 0,-1))\ldots (1, 24-d_1, 0,-1)$. However, the number of
non-toric deformations is precisely equal to the number of missing points.

Thus, even though the polyhedra look different, the manifolds are actually the
same\footnote[1]{We are grateful to Paul Aspinwall for explaining this point
to us.}.
This is because we have ignored the non-toric deformations which are not
visible as points in the polyhedra. In many cases, it is possible to ignore
them because they are often zero.
However, these non-toric deformations account for the observed discrepancy - 
the polyhedra obtained by unhiggsing $E_8$ have precisely as many non-toric
deformations as missing tensor multiplets. Taking these into account,
we are led to conclude that these polyhedra all describe the same
manifold corresponding to a heterotic vacuum with 13 tensor multiplets and 
$E_8$ gauge symmetry.

In the same way we can construct duals to heterotic vacua 
with the first $E_8$ unhiggsed. This amounts to replacing $u$ 
in~(\ref{sc:si}) by $v$ and writing~(\ref{eq:si}) with appropriate power of
the new coordinate.

We can also unhiggs both $E_8$'s which will result
in the unbroken $E_8\times E_8$ visible in the $K3$ subpolyhedron.
The Hodge numbers of our Calabi-Yau are invariably $(43,43)$ and there are 25
tensor multiplets, but not all of them are visible in the fan of the base,
some being accounted for by the non-toric deformations.

\section{Conclusion}
In this paper we have considered F-theory duals of heterotic $E_8\times E_8$
compactifications on $K3$ with instanton numbers $(d_1,d_2)$, and $n_T$ 
tensor multiplets. The duals of these theories are conjectured to be
F-theory compactifications on elliptic Calabi-Yau threefolds 
\cite{V:F,MV2}. This
duality is related to the heterotic/type IIA dualities 
proposed in \cite{KV,AFIQ}.
It was observed in \cite{poly:CF} that the sequences of reflexive polyhedra 
associated to these spaces are nested so as to reflect heterotic perturbative 
and non-perturbative processes. Note that the fact that these polyhedra are
nested implies that the moduli spaces of the corresponding Calabi-Yau
manifolds are connected \cite{Conn}. 
The new contribution here is to show that the
sequences of reflexive polyhedra in \cite{poly:CF} can 
be obtained systematically
using hypersurfaces in toric varieties, which are blow-ups of the
{\rm I\kern-.18em F}$_{n}$ 
models described in \cite{MV2}. We also find that the blowups are encoded
torically by extra points in the fan of the {\rm I\kern-.18em F}$_{n}$, which
is visible in the dual polyhedra.  
 
The models that we find would be the lowest members of 
chains of duals which can
be obtained by unhiggsing in the spirit of refs.
\cite{BKKM,AFIQ,poly:CF,Ber}. 

We also show how to construct Calabi-Yau's corresponding to heterotic models
by unhiggsing either, or both, $E_8$'s. We find that in such cases there are
as many extra tensor multiplets as there were instantons in the unhiggsed
$E_8$, but not all of them appear as points corresponding to blowups of the 
fan of the {\rm I\kern-.18em F}$_{n}$ in the dual polyhedra - some are
encoded non-torically.

\bigskip
{\Large\bf Acknowledgements}

We would like to thank A. Font and J. Distler for useful discussions. We would
also like to thank P. S. Aspinwall for pointing out an error in an earlier 
version.This work was supported in part by the Robert A. Welch Foundation and
the NSF grant PHY-9511632.

\pagebreak

\end{document}